\documentclass[aps,nofootinbib,notitlepage,longbibliography,twocolumn, superscriptaddress]{revtex4-1}
\usepackage{amsmath,amssymb}
\usepackage{slashed}
\usepackage{graphicx,amsfonts}
\usepackage{bm}
\usepackage[T1]{fontenc}
\usepackage[utf8]{inputenc}
\usepackage{gauss} 
\usepackage[top=1.0in,bottom=1.0in,left=1.0in,right=1.0in]{geometry}
\usepackage[colorlinks,linkcolor=blue,citecolor=blue]{hyperref}
\usepackage{subfig}

\newcommand{\be}{\begin{equation}}
\newcommand{\ee}{\end{equation}}
\newcommand{\bea}{\begin{eqnarray}}
\newcommand{\eea}{\end{eqnarray}}
\newcommand{\ba}{\begin{eqnarray}}
\newcommand{\ea}{\end{eqnarray}}

\def\be{\begin{eqnarray}}
\def\ee{\end{eqnarray}}
\def\bea{\be}
\def\eea{\ee}

\def\roughly#1{\mathrel{\raise.3ex\hbox{$#1$\kern-.75em%
\lower1ex\hbox{$\sim$}}}}

\usepackage{lipsum}

\newcommand{\dd}{\mathrm{d}}

\date{\today}
\begin{abstract}
 We analyze the entanglement of a Schwinger pair created by a time-dependent pulse. In the semiclassical approximation,  the pair creation by a pulse of external electric field is captured by a periodic worldline instanton. At strong gauge coupling, the gauge-gravity dual worldsheet instanton exhibits a falling wormhole in AdS. We identify the tunneling time at the boundary with the inverse Unruh temperature, and derive the pertinent entanglement entropy between the created pair using thermodynamics. The entanglement entropy is enhanced by the sub-barrier tunneling process, and partly depleted by the radiation in the post-barrier process.
\end{abstract}
\begin{document}
\title{Entanglement entropy in a time-dependent holographic Schwinger pair creation}
\author{Sebastian Grieninger}
\email{sebastian.grieninger@stonybrook.edu}
\affiliation{Center for Nuclear Theory, Department of Physics and Astronomy,
Stony Brook University, Stony Brook, New York 11794–3800, USA}

\author{Dmitri E. Kharzeev}
\email{dmitri.kharzeev@stonybrook.edu}
\affiliation{Center for Nuclear Theory, Department of Physics and Astronomy,
Stony Brook University, Stony Brook, New York 11794–3800, USA}
\affiliation{Department of Physics, Brookhaven National Laboratory
Upton, New York 11973-5000, USA}

\author{Ismail Zahed}
\email{ismail.zahed@stonybrook.edu}
\affiliation{Center for Nuclear Theory, Department of Physics and Astronomy,
Stony Brook University, Stony Brook, New York 11794–3800, USA}
\maketitle

\section{Introduction}
Schwinger pair creation~\cite{Schwinger:1951nm} is a
quantum process in which a pair of particles is produced from the vacuum in  an external  electric field. In quantum chromodynamics (QCD), the external chromoelectric field inside a confining string can lead to the production of a quark-antiquark  pair, leading to string breaking  when the string energy exceeds the light meson mass. This process is used in event generators (such as Lund and  PYTHIA)~\cite{Andersson:1983ia}  to account for jet fragmentation in high energy collisions ~\cite{ALEPH:1995qic}. 

The pair creation proceeds through tunneling, a quintessential quantum process. The vacuum pair tunnels through a potential barrier under the effect of a strong background
field, and is produced with a finite probability. Throughout the creation process, the pair is correlated and interacts with both with the
external (classical) field and the quantum gauge field. Quantum entanglement is the measure of these correlations  throughout the pair history.

Recently, this pair creation process was used as an illustration of the Einstein-Rosen-Podolsky (EPR) paradox, in the context of  quantum field theory at strong coupling~\cite{Jensen:2013ora,Sonner:2013mba}. In this strong coupling regime, the gravity dual of the pair is a string worldsheet~\cite{Gorsky:2001up,Xiao:2008nr,Semenoff:2011ng,Lewkowycz:2013laa,Chernicoff:2013iga,Jensen:2014bpa,Hubeny:2014zna,Ghodrati:2015rta,Semenoff:2018ffq,Yeh:2023avs} with a non-traversable wormhole, or Einstein-Rosen (ER) bridge~\cite{Jensen:2013ora}. The quantum entanglement entropy of the pair is sensitive to the location of the ER bridge~\cite{Jensen:2013ora,Maldacena:2013xja}.

In the receding quark-antiquark pair in Schwinger process, the end-points of the string are never causally connected.
However the pair is entangled owing to its color neutrality. In non-confining dual gravity
description, the entanglement entropy was found to be of order $\sqrt{\lambda}$ in the weak field limit~\cite{Jensen:2013ora,Hubeny:2014zna,Hubeny:2014zna,Grieninger:2023ehb,Hubeny:2014kma}, with $\lambda=g_Y^2N_c$ the strong $^\prime$t Hooft coupling. In this work, we will extend our discussion which was restricted to static electric fields \cite{Grieninger:2023ehb} and explore the entanglement in the pair creation
process in the presence of a time-dependent electric pulse.

In QCD, a strong gauge field pulse could be generated by a highly boosted nucleus, assuming that all
wee partons in the nuclear wavefunction add up coherently, see ~\cite{Gelis:2010nm,Kovchegov:2012mbw} and references therein. In the rest frame of a target interacting with this highly boosted nucleus, the target is crossed by a ``gluon wall" \cite{bjorken1992black} that is a pulse of gauge field. This pulse will lead to pair creation. The entanglement entropy of the pair can be converted to the Gibbs entropy of the final hadronic state. For QED, the equality between the entanglement and Gibbs entropies in Schwinger pair production by a pulse was demonstrated in \cite{florio2021gibbs}. For CFTs it was shown in \cite{Casini:2011kv} that the entanglement entropy across a spherical region is equal to the thermal entropy of the hyperbolic geometry $\mathbb R \times\mathbb{H}^{d-1}$, which is the direct product of time and the
hyperbolic plane. In QCD, the effects of confinement have to be taken into account compared to the weak coupling QED calculation in \cite{florio2021gibbs}, as well as radiation in the final state. Our present work addresses this problem from a holographic perspective.

The organization of the paper is as follows:  In section~\ref{SECA} we briefly review the semiclassical analysis of the Schwinger process for particle pair creation by a pulse. The worldline instanton solution is discussed, and an estimate of its entanglement entropy is given. In section~\ref{SECB}, we extend the analysis to strong coupling, where the dual of the worldline is a worldsheet in AdS. The worldsheet is characterized by a moving wormhole, a hallmark of the time-varying pulse at the boundary. We show that the entanglement entropy receives a positive contribution from  tunneling (sub-barrier process) and a negative contribution from   radiation loss (post-barrier process). Our conclusions are in section~\ref{SECC}.

\section{Tunneling in a time-dependent external field}
\label{SECA}
Consider first  a scalar particle of mass $M$, in an Abelianized
external field moving in proper time, with the action
\bea
\label{A1}
S=\int d\tau M\sqrt{\dot{x}^2}+i\oint A.
\eea
Tunneling in (\ref{A1}) is captured by the worldline instanton solution to the classical equations of motion in Euclidean signature,
\bea
\label{FORCE}
M\frac{d}{d\tau}\bigg(\frac{{\dot{x}_\mu}}{\sqrt{\dot{x}^2}}\bigg)=iF_{\mu\nu}\dot{x}_\nu.
\eea
For a time dependent pulse electric field in Minkowski signature,
\bea
\label{PULSE}
E(t)
%=E\,{\rm sech^2}(\omega t)
=\frac{E}{{\rm cosh}^2(\omega t)}
\eea
 the Euclideanized vector potential in (\ref{A1})
\bea
\label{A3X}
A_3(x_4)=\frac {-iE}{\omega}{\rm tan}(\omega x_4)
\eea
is purely imaginary.  Since (\ref{A3X}) is singular for $x_{4S}=\frac \pi{2\omega}$, all worldlines
are bounded by this maximum value in $|x_4|\leq x_{4S}$.

\begin{figure}
    \centering
    \includegraphics[width=0.6\linewidth]{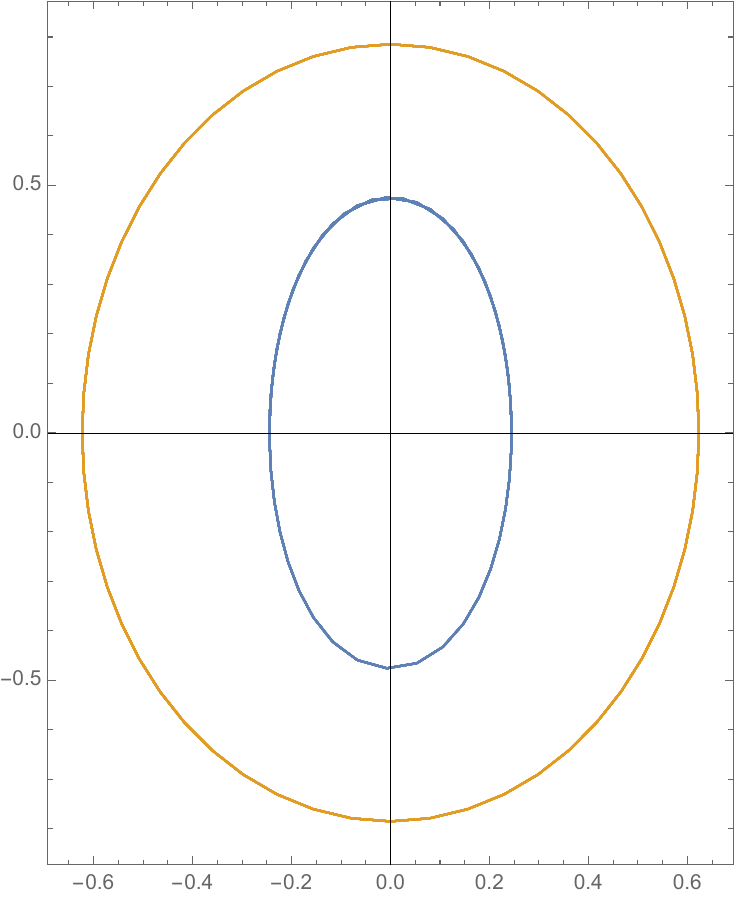} 
    \caption{Contour plot of (\ref{CONTOUR12}) for $a=\gamma=1$ (outer) and $a=1, \gamma=2.5$ (inner) with $M=1$.}
    \label{fig:cont}
\end{figure}

\subsection{Worldline instanton}
The tunneling process was discussed by many~\cite{Brezin:1970xf,Popov:1971iga,Kim:2000un,Dunne:2005sx,kharzeev2005color,Gies:2005bz,Gies:2015hia,Gies:2016coz}, with the instanton solution  given by~\cite{Dunne:2005sx}
\begin{widetext}
\bea
\label{W1}
&&x_3(\tau)=\frac 1{a\gamma} \frac 1{\sqrt{1+{\gamma}^2}}
{\rm arcsinh}\bigg(\gamma{\rm cos}\bigg(a\sqrt{1+\gamma^2}\tau\bigg)\bigg)
\equiv \frac 1{a\bar\gamma}{\rm sinh}^{-1}(\gamma{\rm cos}(\bar a\tau)),
\nonumber\\
&&x_4(\tau)=\frac 1{a\gamma}
{\rm arcsin}\bigg(\frac{\gamma}{\sqrt{1+{\gamma}^2}}{\rm sin}\bigg(a\sqrt{1+\gamma^2}\tau\bigg)\bigg)\,\,\,\,\,\,\,\,
\equiv \frac 1{a\gamma}{\rm sin}^{-1}(\underline\gamma{\rm sin}(\bar a \tau)).
\eea
\end{widetext}
It is periodic, with period 
\bea
\label{PERIODX}
\beta=\frac {2\pi}{\bar a}\equiv \frac{2\pi}{a\sqrt{1+\gamma^2}}
\eea
with $a=E/M$ and  $\gamma=\omega/a$.
Throughout, $E\equiv g_YE$ refers to the invariant electric field in the probe D3 brane. Eqs (\ref{W1}) describe a cyclotron-like  trajectory that satisfies
$\dot{x}_3^2+\dot{x}_4^2=1$,
since the applied force in (\ref{FORCE}) is magnetic in Euclidean signature.
The Wilson loop traced by the boundary worldline (\ref{W1}) is elliptic-like in general,
\bea
\label{CONTOUR12}
\frac{{\rm sh^2}(a\bar\gamma x_3)}{\gamma^2}+\frac{{\rm sin^2}(a\gamma  x_4)}{\underline\gamma^2}=1
\eea
as illustrated in Fig.~\ref{fig:cont}. The high eccentricity of the trajectories along the $x_4$
direction follows from the singularity of the force from the pulse,  at $x_{4S}=\frac \pi{2\omega}$ of (\ref{A1}) noted earlier. It  becomes circular in the static limit, $x_3^2+x_4^2\rightarrow \frac 1{a^2}$ as 
$\omega\rightarrow 0$.

\subsection{Particle entanglement}
For scalar pair creation of mass $2M$, the action evaluated using (\ref{W1}) is
\bea
\label{W4}
S(\beta)=\frac{\beta M}{1+(1-(\beta\omega/2\pi)^2)^{\frac 12}}
\eea
with  $e^{-2S(\beta)}$, the penalty for tunneling in the absence of confinement. 
When the pulse is about static, $S(\beta)\sim \frac 12 \beta M$ in agreement with
Schwinger$^\prime$ original result. Alternatively, when the pulse is very sharp 
in time, $S(\beta)\sim \frac{2\pi M}\omega$.  At large frequency, the pair production
is uninhibited.

We interpret (\ref{W4}) as the free energy $S(\beta)=\beta F(\beta)$
of the tunneling pair in the pulse. As a result, the corresponding  quantum entropy at the boundary, is given by thermodynamics
\bea
\label{W5}
S^B_{EE}=&&\frac{\beta M\gamma^2}{(1+\sqrt{1+\gamma^2})^2}.
\eea
It vanishes in the static limit  as $S_{EE}^B\sim \beta M\gamma^2$, and for large frequencies 
vanishes as  $S_{EE}^B\sim\frac {2\pi M}\omega$.

\begin{figure*}
    \centering
    \includegraphics[width=0.35\linewidth]{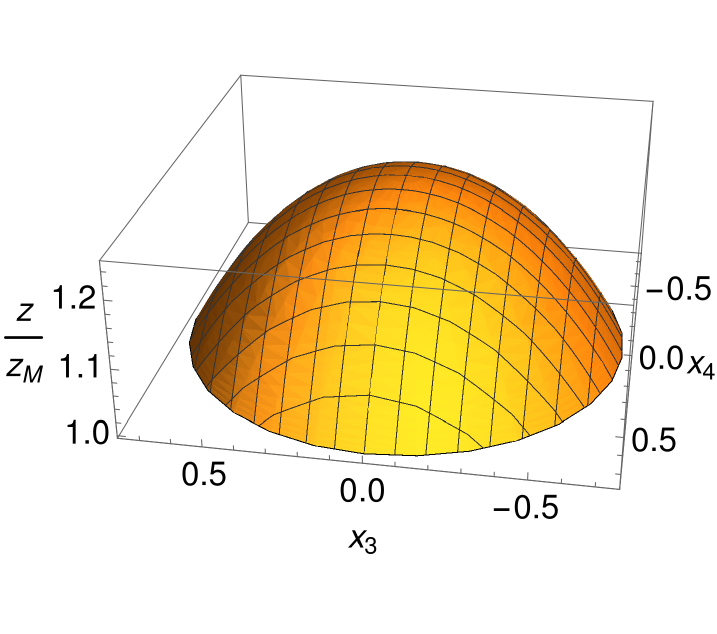}
     \includegraphics[width=0.40\linewidth]{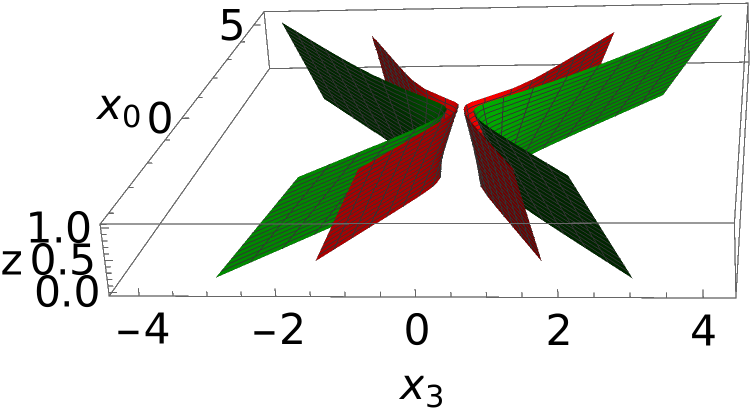} 
    \caption{Left: Sub-barrier Euclidean string worldsheet, for pair production in a pulse for $z_ME=0.44$ and $z_M\omega=0.25$. Right:
    Post-barrier Minkowski string worldsheet, for pair production in a pulse, for $\gamma=a=1$ (green) and $\gamma=2, a=1$ (red), with $z_M=0$.}
    \label{fig:numsurface}
\end{figure*}

\section{Holographic pair production}
\label{SECB}
The worldline instanton (\ref{W1}) captures the pair production of a particle of mass $M$ at the boundary, at weak coupling. In the double limit  of strong $^\prime$t Hooft coupling $\lambda$ and large $N_c$, the gravity dual description is captured by a string worldsheet sourced
by the boundary Wilson loop on a D3 brane. Holographic pair production of particles as end points of strings, extends (\ref{A1}) to the Nambu-Goto action in bulk
\bea
\label{A1S}
S=\sigma_T\int \dd\tau\, \dd\sigma |{\rm det} g_{MN}\,\partial_\tau X^M\partial_\sigma X^N|^{\frac 12} +i\oint A\nonumber\\
\eea
in AdS$_5$ with line element
\bea
\dd s^2=g_{MN}\dd x^M\dd x^N=\frac{L^2}{z^2}(\dd x_\mu^2+\dd z^2)
\eea
with $\sigma_TL^2=\sqrt\lambda/2\pi$. The coordinate embedding of the string $X^M(\tau, \sigma)$
can be obtained numerically in Euclidean signature, using the Nambu-Goto action for the string in bulk, as constrained by the Wilson loop at the D3 boundary. 

\begin{figure*}
    \centering
 \includegraphics[width=0.48\linewidth]{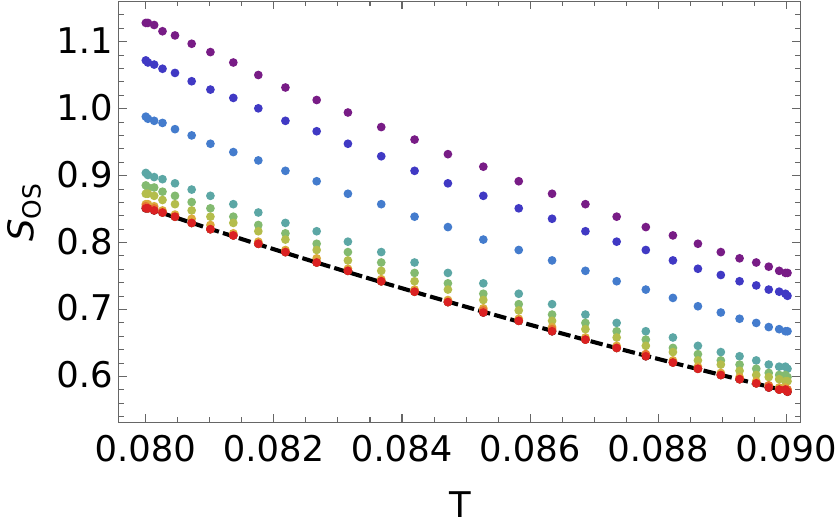}
  \includegraphics[width=0.48\linewidth]{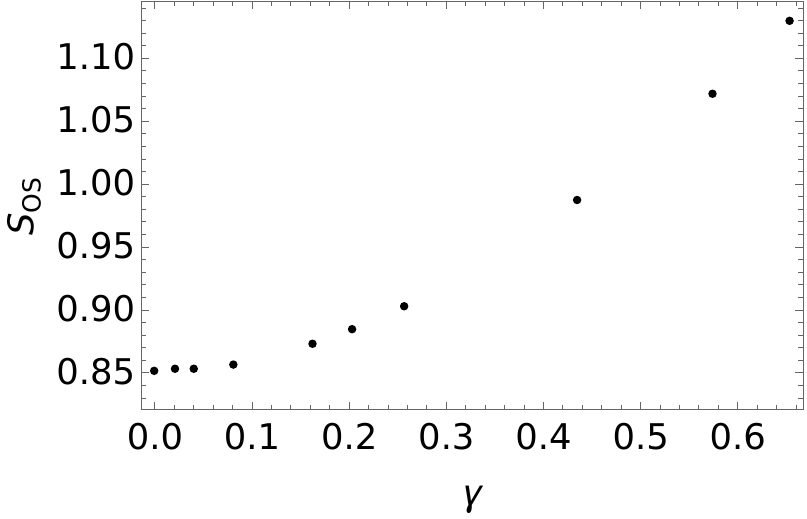}
    \caption{Left: On-shell action vs $T=1/\beta$ for different $
    \omega$ (increasing from red to purple). The analytical result for $\omega=0$ is the black dashed line. Right: On-shell action vs $\gamma$ for $T=0.08$ and $\lambda=12$.}
    \label{fig:osnumerics}
\end{figure*}

\subsection{Euclidean string worldsheet}
\label{SECAPP}
In Euclidean signature, the  worldsheet surface in bulk is ellipsoidal.
To parametrize it in AdS$_5$, we use cylindrical coordinates $r,\varphi, z$,
\bea
\label{WX1}
x^{M}\!(z, \varphi)\!=\!(0,0,r(z,\varphi){\rm cos}(\varphi), r(z,\varphi){\rm sin}(\varphi), z)
\eea
in terms of which the Nambu-Goto action plus the electric field at the boundary,  read
\begin{widetext}
\bea
\label{WX2}
    S=\frac{\sqrt{\lambda}}{2\pi}\int_{z_M}^{z_{E}}
    \dd z\int_0^{2\pi} \dd \varphi\, \frac 1{z^2}\bigg(r^2(1+r_z^2)+r_{\varphi}^2\bigg)^{\frac 12}
    +i\oint A
\eea
\end{widetext}
with the shorthand notation $r_z=\partial_zr$ and $r_\varphi=\partial_{\varphi}r$. Here $z_M$ is the position of the probe D3 brane, with $z_M=\frac{\sqrt\lambda}{2\pi M}$
 fixed by the mass of the probe D3 brane at the boundary.  Here
$z_E$ is the tip (extrema) of the ellipsoidal-like worldsheet  in bulk,  
\bea
\label{WX3}
r(z_E,\varphi)=0\qquad r_z(z_E, \varphi)=\infty.
\eea
In  (\ref{WX2})  we have used the fact that the electric field (\ref{W1}) acts as a magnetic field in Euclidean signature, with the "vector potential" (\ref{A3X}). Its  contribution to the Wilson loop at the boundary $z=z_M$, is
\bea
i\oint A=
\int_0^{2\pi}\dd \varphi \frac E\omega {\rm tan}(\omega r{\rm sin}\varphi)
({\rm cos}\varphi r_\varphi-r{\rm sin}\varphi).\nonumber\\
\eea

The bulk equation of motion follows by variation
\bea
\label{WX4}
&&-r r_{zz}(r^2+r_\varphi^2)+rr_{\varphi\varphi}(1+r_z^2)\nonumber\\
&&-\frac 2z r_\varphi(r^2(1+r_z^2)+r_\varphi^2)\nonumber\\
&&-r^2r_z^2-2r_{\varphi}^2-2rr_zr_\varphi r_{z\varphi}-r^2=0
\eea
subject to the Lorentz force along the z-direction at the D3 brane on the  boundary at $z=z_M$,
\bea
&&\frac{\sqrt\lambda}{2\pi}\frac 1{z_M^2}\frac {r^2r_z}{(r^2(1+r_z^2)+r_\varphi^2)^{\frac 12}} \nonumber\\
&&= 
\label{WX6}
\frac E{{\rm cos}^2(\omega r{\rm sin}\varphi)}\bigg(r-\frac 12 r_\varphi\, {\rm sin}{2\varphi}\bigg).%\nonumber\\
\eea
Again, we made use of the  short hand notations $r_{zz}=\partial_z^2 r$,
$r_{\varphi\varphi}=\partial_\varphi^2 r$, and $r_{z\varphi}=\partial_z\partial_\varphi  r$.
\eqref{WX4} subject to the boundary condition \eqref{WX6} can be solved numerically using the pseudo-spectral methods (with Chebychev discretization in the radial direction and Fourier discretization in $\varphi$).

In Fig.~\ref{fig:numsurface} (left) we  show the Euclidean surface for $z_ME=0.44$ and $z_M\omega=0.25$. The on-shell action $S_{OS}$ of the ellipsoidal surface (\ref{WX2}),  sets the tunneling probability for the holographic pair production in the pulse, in the strong field limit. In  Fig.~\ref{fig:osnumerics} (left) we show the behavior of the on-shell action versus $T=1/\beta$ given in (\ref{PERIODX}), for increasing frequencies of the pulse from bottom-red to top-purple. The numerical results shown in the dashed-black curve for $\omega=0$, coincide with the analytical results for the
holographic pair production derived in~\cite{Semenoff:2011ng}
\bea
\label{ZERO}
S_{OS}\rightarrow \frac{\sqrt\lambda}2
\bigg(\sqrt{\frac{T_c}{T}}-\sqrt{\frac{T}{T_c}}\bigg)^2
\eea
with 
$$T_c=\frac {a_c}{2\pi}=\frac 1{2\pi}\frac 1{\sqrt{R^2+z_M^2}}.$$
Our numerical results, generalize (\ref{ZERO}) to a time-dependent pulse.

In  Fig.~\ref{fig:osnumerics} (right) we show  the on-shell action versus $\gamma=M\omega/E$ for fixed temperature. The increasing action reflects on the larger penalty for the pair production rate, with larger $\omega$ and fixed $E$ as the pulse is short lived.  Equivalently, the larger $E$ for fixed $\omega$, the smaller the penalty for pair production.% More specifically, 

\subsection{Particle multiplicities in a pulse}
Strong and coherent chromo-electric pulses can be
produced during the initial phase of 
an ultra-relativistic heavy-ion collision.
The higher the energy of the ion projectile, the stronger the field and the shorter is its duration in the frame of the target. 

\begin{figure}
    \centering
    \includegraphics[width=1.0\linewidth]{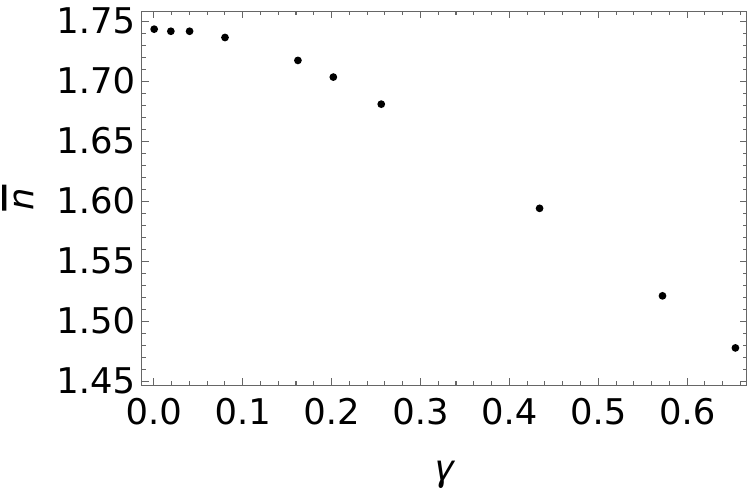} 
    \caption{Mean number of particles produced in the pulse as a function of $\gamma=M\frac\omega E$ for $T=0.08$, $z_M=1$ and $\lambda=12$.}
    \label{fig:meann}
\end{figure}

Strong chromo-electric fields can produce light  quark pairs through  the Schwinger mechanism. For the strongly coupled phase, the mean number of produced pairs in the pulse is 
\bea
\bar{n}\!=\!\sum_{n=1}^\infty n\,p_n\!=\!
\sum_{n=1}^\infty n\,(e^{S_{OS}}\!-1)\,e^{-nS_{OS}}\!=\!
\frac{1}{1-e^{-S_{OS}}}.\nonumber
\eea
In Fig.~\ref{fig:meann}, we show the dependence of the mean number of produced particles on $\gamma=M\frac \omega E$
(not to be confused with the Lorentz contraction factor). The mean number of produced pairs decreases with increasing $\gamma$, since tunnelling is more suppressed (note that the case of strong and static electric fields corresponds to $\gamma \to 0$.)

\subsection{Minkowski string worldsheet}
In Minkowski signature, the worldsheet in bulk is the locus of the retarded radiation,
sourced by an accelerating $q\bar q$ pair tracing the Wilson loop with hyperbolic-like worldlines. It follows from the ruled surface in bulk~\cite{Mikhailov:2003er}
\bea
\label{R1}
X^M(t, z)=(z\dot{x}^\mu (t)+x^\mu(t), z)
\eea
with $\sigma=z$. The boundary worldlines 
$$x^\mu(t)=(x^0(t),0,0,x^3(t))$$ 
follow from (\ref{W1}) by analytical continuation
$\tau\rightarrow it$ and $ix_4\rightarrow x_0$, 
\bea
\label{W1M}
&&x^3(t)=\frac 1{a\bar\gamma}{\rm sinh}^{-1}(\gamma{\rm cosh}(\bar at))
\nonumber\\
&&x^0(t)= \frac 1{a\gamma}{\rm sinh}^{-1}(\underline\gamma{\rm sinh}(\bar a t))
\eea
with $\dot{x}^2_\mu= -1$. In Fig.~\ref{fig:numsurface} (right) we show two worldsheets as given by  (\ref{R1}) and traced by (\ref{W1M}) with $\gamma=a=1$ (green) and $\gamma =2, a=1$ (red). We have set $z_M=0$ for the D3 brane for convenience. For $\gamma=0$, the ruled surface (right) is the analytical continuation of the Euclidean surface (left), as initially observed in~\cite{Jensen:2013ora,Sonner:2013mba}.

\subsection{Moving wormhole on the worldsheet}
The time-dependent string worldsheet harbors  a moving wormhole.  To see this, consider the line element associated to (\ref{R1}) 
\bea
\dd X_M^2=\left(a^2(t) -\frac 1{z^2}\right)\dd t^2-\frac 2{z^2}\dd t\, \dd z
\eea
with a squared  acceleration
\bea
\label{ACC2}
a^2(t)\equiv \ddot{x}^2= a^2
\bigg(1+\underline\gamma^2{\rm sinh}^2(\bar a t)\bigg)^{-2}
\eea
and an effective horizon at $z_H(t)=1/a(t)$.
In the static limit $\omega\rightarrow 0$, it reduces to the static horizon $z_H\rightarrow \frac 1a$. Away from the static limit, the horizon is moving away from the boundary,
with asymptotically
\bea
\label{ASYMP}
z_H(t)\sim \frac{\gamma\underline\gamma}{4{\overline a}}\,e^{2\overline a t}.
\eea
The black hole is falling rapidly  in bulk, a hallmark of a time-dependent problem~\cite{Shuryak:2005ia,Kim:2007ut,Grieninger:2022yps}.

\section{Entanglement for the pulsing string}
The effective horizon $z_H(t)$ splits the worldsheet in bulk into a causal 
part with $z<z_H(t)$ and a  non-causal part with $z>z_H(t)$. This is also the location of a time-dependent wormhole, as observed in~\cite{Jensen:2013ora} for the static case.

The entanglement entropy (EE) receives contribution from both the causal part of the worldsheet (positive)
and non-causal part of the worldsheet (negative). We will distinguish between the weak field limit and strong coupling where the worldsheet surface can be obtained both analytically and numerically (for the causal part only), and the strong field and strong coupling limit where
the worldsheet surface can only be obtained numerically.

\begin{figure*}
    \centering
 \includegraphics[width=0.48\linewidth]{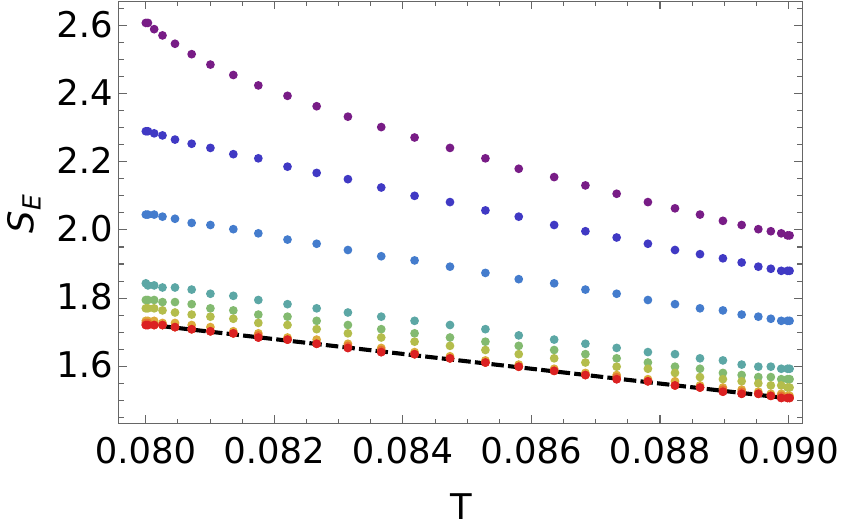}
    \includegraphics[width=0.48\linewidth]{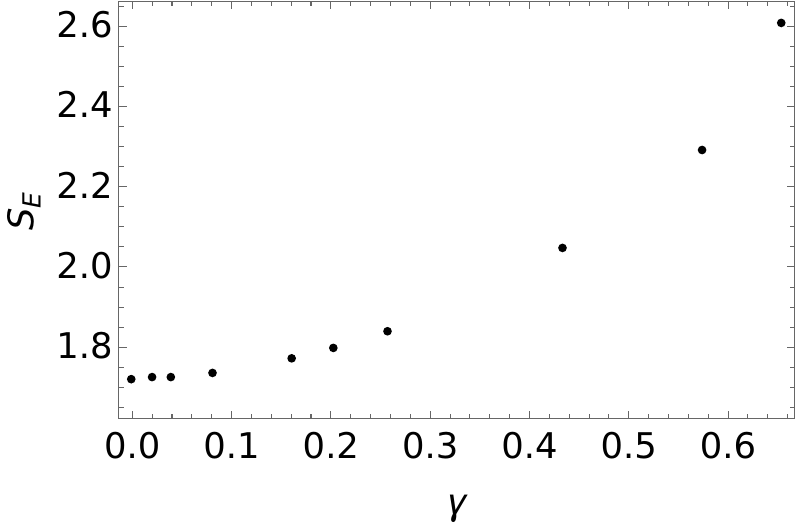}
  \caption{Left: Entanglement entropy as a function of the temperature $T=1/\beta$  for different $
    \omega$ (increasing from red to purple). The analytical result for $\omega=0$ is the black dashed line. Right: Entanglement entropy as a function of $\gamma=\omega/a$ for $T=0.08$, $z_M=1$ and $\lambda=12$.}
    \label{fig:EEnumerics}
\end{figure*}

\subsection{Causal contribution to EE: \\Weak field limit}
The causal contribution to the entanglement entropy in the weak field limit and finite $\gamma=M\omega/E$ is given numerically by the Euclidean surface, and more explicitly by
the ruled surface in Minkowski signature, as we have discussed in~\cite{Grieninger:2023ehb}. Specifically, 
the causal contribution of the worldsheet action
\bea
S^C=&&2\times\frac{\sqrt\lambda}{2\pi}\int_{-\frac 12 {\cal T}}^{+\frac 12 {\cal T}}\dd t
\int_{z_M}^{\frac 1{a(t)}} \frac {\dd z}{z^2}\nonumber\\
=&&2\times\frac{\sqrt\lambda}{2\pi}\int_{-\frac 12 {\cal T}}^{+\frac 12 {\cal T}}\dd t
\bigg(\frac 1{z_M}-a(t)\bigg).\label{eq:ncS}
\eea
The first contribution (from the minuend in \eqref{eq:ncS}) reduces to the rest mass contribution
\bea
S_{(1)}^C=\frac{\sqrt\lambda}{\pi\,z_M}\int_{-\frac 12 {\cal T}}^{+\frac 12 {\cal T}}\!\!\dd t=
2\times\,{\cal T}\,\frac{\sqrt\lambda}{2\pi z_M}=2{\cal T} M.
\eea
For $\omega\neq 0$, the time integration of the subtracted term in \eqref{eq:ncS} gives
\begin{widetext}
\begin{align}
S_{(2)}^C=-\frac{\sqrt\lambda}{\pi}\int_{-\frac 12 {\cal T}}^{+\frac 12 {\cal T}}\dd t\,
a(t)=2\times\frac{\sqrt\lambda}{2\pi}\,{\rm ln}\left|\frac{\left(\sqrt{1+\gamma^2}+1\right) e^{-\bar a {\cal T}}+\sqrt{1+\gamma^2}-1}{\left(\sqrt{1+\gamma^2}-1\right) e^{-\bar a {\cal T}}+\sqrt{1+\gamma^2}+1}\right|
\label{eq:seccontr}
\end{align}
\end{widetext}
after using (\ref{ACC2}).

In the large ${\cal T}$ limit, \eqref{eq:seccontr} reduces to
\begin{equation}\label{eq:largeTsecond}
    S_{(2)}^C(\mathcal T\to\infty)=2\times\frac{\sqrt\lambda}{2\pi}\,{\rm ln}\bigg|\frac{\sqrt{1+\gamma^2}-1}{\sqrt{1+\gamma^2}+1}\bigg|.
\end{equation}
In the small frequency limit with $|\gamma|\ll 1$, (\ref{eq:largeTsecond}) reduces to $-\frac{2\sqrt\lambda}\pi{\rm ln}|\frac 1\gamma|$, which is to be compared to~\cite{Grieninger:2023ehb} 
\bea
\label{ZERO}
2M{\cal T}-2\sqrt{\lambda}{\cal T}/\beta
\eea
in
the static limit. This limit is singular and does not reduce to our static result~\cite{Grieninger:2023ehb} . We conclude that the small frequency limit does not commute with the large time limit.
If we take the small frequency limit first, then \eqref{eq:seccontr} reduces to\begin{widetext}
 \begin{equation}\label{eq:smallfreq}
    S_{(2)}^C(\omega\to0)=-2\sqrt{\lambda} T\,\mathcal T+\sqrt{\lambda}\,\frac{\sinh(2\pi T\,\mathcal T)}{8\pi^3T}\,\omega^2+\mathcal O(\omega^4).
\end{equation}   
\end{widetext}
Eq. \eqref{eq:smallfreq} correctly reproduces the static case considered in~\cite{Grieninger:2023ehb}.
In the following, we assume that $\omega$ is sufficiently large. Recall, that we are in the weak field limit, i.e. $E=M\cdot a\ll 1$. In the weak field limit, \eqref{eq:seccontr} reduces to\begin{widetext}
 \begin{equation}\label{eq:smallasecond}
    S_{(2)}^C(a\to0)=-\frac{2 \sqrt{\lambda } \sqrt{4 \pi ^2 \,T^2-\omega ^2} \tanh \left(\frac{\mathcal T \omega }{2}\right)}{\pi  \omega }+\mathcal O(a^3).
\end{equation}   
\end{widetext}
If $\omega$ is not too small, we can safely take the large $\mathcal T$ limit with  $\tanh (\mathcal T \omega /2)\to 1$. Hence, $S^C_{(2)}$ is finite in the large $\mathcal T$ limit and thus subleading.

With this in mind  and using (\ref{ZERO}),  the 
causal contribution is
\bea
S^C= 2{\cal T} M -2\sqrt{\lambda}{\cal T}/\beta (1-\theta(|\gamma|))
+{\cal O}({{\cal T}^0})
\eea
This is  expected for $\gamma\neq 0$, since the effective horizon asymptotes the Poincar\'e singularity 
for large times as we noted in (\ref{ASYMP}). At late times, the self-energy is solely due to the 
mass following from the D3 brane, with no Debye screening mass induced by the rapid fall off. We now interpret the causal part of the action $F^C=S^C/\beta$  as a {\it free energy} for fixed temperature $T=1/\beta$~\cite{Grieninger:2023ehb}. In the weak field limit, 
 the causal contribution of the EE for the pulse is then identified through thermodynamics
\bea
\label{SEEC2}
S_{EE}^C(\gamma)= \beta^2\frac{\partial F^C}{\partial\beta}=\sqrt\lambda \,(1-\theta(|\gamma|))+{\cal O}({{\cal T}^0}),
\eea
which is equal to zero at late times for finite $|\gamma|$.

\subsection{Causal contribution to EE:\\ strong field limit}
The causal contribution to the entanglement entropy in the strong field limit, is solely given by the numerically generated Euclidean worldsheet, using the arguments we presented in~\cite{Grieninger:2023ehb}. Again, we can regard the on-shell action $S_{OS}=\beta F_{OS}$ as a function of the temperature $T=1/\beta$ shown
in Fig.~\ref{fig:osnumerics} (left),  as a {\it free energy} $F_{OS}$. In the strong field limit, the  EE is again  identified through thermodynamics
\bea
\label{SEENUM}
S^C_{E}=\beta^2\frac{\partial F_{OS}}{\partial\beta}
\eea

In Fig~\ref{fig:EEnumerics} (left) we show the numerical results for 
the EE as given by (\ref{SEENUM} versus temperature $T=1/\beta$, for
different pulse frequencies $\omega$. The curves are for increasing frequencies from bottom-red to top-purple. The black-dashed curve
is the exact result~\cite{Grieninger:2023ehb} for $\omega=0$.

In the right side of Fig~\ref{fig:EEnumerics}, we show the EE versus $\gamma=\omega/a$. In the pulsing electric field, the EE is 
continuously enhanced.

\subsection{Non-causal contribution to EE:\\ Weak field limit}
The non-causal part of the entanglement entropy, amounts to evaluating the radiation loss across the
falling horizon. This is readily done by noting that (\ref{R1}) with the time-dependent acceleration (\ref{ACC2}), captures the Larmor radiation at strong  coupling~\cite{Mikhailov:2003er}
\begin{widetext}
\begin{align}
\label{LARMOR}
&{\cal E}_R=\frac{\sqrt\lambda}{2\pi}\int^{+\frac 12{\cal T}}_{-\frac 12{\cal T}}\!\!\! \dd t\, a^2(t)=\frac{a \sqrt{\lambda }}{4 \pi } \!\left(\!-\left(2+\gamma ^2\right) {\rm ln}\left|\frac{\left(\sqrt{1+\gamma^2}+1\right) e^{-\bar a {\cal T}}\!+\!\sqrt{1+\gamma^2}-1}{\left(\sqrt{1+\gamma^2}-1\right) e^{-\bar a {\cal T}}\!+\!\sqrt{1+\gamma^2}+1}\right|-\frac{\gamma ^2 \sqrt{1+\gamma^2} \left(e^{\bar a{\cal T}}-e^{-\bar a {\cal T}}\right)}{\frac{\gamma ^2}{2}  \left(e^{-\bar a {\cal T}}+e^{\bar a  {\cal T}}\right)+\gamma ^2+2}\right)\nonumber\\&
\end{align}
\end{widetext}
which simplifies in the large time limit to
\begin{equation*}
-\frac{a \sqrt{\lambda }}{4 \pi } \left(2 \sqrt{\gamma ^2+1}+\left(\gamma ^2+2\right) {\rm ln}\! \left|\frac{\sqrt{1+\gamma ^2}-1}{\sqrt{1+\gamma^2}+1}\right|\right).\nonumber\\
\end{equation*}

The radiation loss of the entanglement entropy follows by interpreting (\ref{LARMOR}) as free energy and taking the derivative with respect to the temperature.
The temperature follows by using (\ref{PERIODX}) to redefine the acceleration in terms of the effective temperature for
fixed $\omega$ as in (\ref{PERIODX}).
If we identify $\mathcal T=\mathcal N/T$ as the luminal time it takes the radiation to fall to the effective horizon~\cite{Grieninger:2023ehb}, the entanglement entropy follows as $S_{EE}^{NC}=- \partial {\cal E}_R/\partial T$. The constant $\mathcal N$ fixes the time it takes to reach the horizon and was determined in \cite{Mikhailov:2003er,Grieninger:2023ehb} as $\mathcal N=1/(3\pi)$. Hence, we find
\begin{widetext}
\begin{align}\label{SEE2}
&\frac{S_{EE}^{NC}(\gamma)}{\sqrt{\lambda } }=\!\!-\frac{1}{2} \sqrt{1+\gamma ^2} \!\left(\!\frac{2 \sqrt{1+\gamma ^2} \gamma ^2\left(e^{4/3}-1\right) \left(\left(1+e^{2/3}\right)^2 \gamma ^2+12 e^{2/3}\right) }{\left(\left(1+e^{2/3}\right)^2 \gamma ^2+4 e^{2/3}\right)^2}+\left(\gamma ^2-2\right) \log \left(\frac{\sqrt{1+\gamma ^2}-\tanh \left(\frac{1}{3}\right)}{\sqrt{1+\gamma ^2}+\tanh \left(\frac{1}{3}\right)}\right)\right)
\end{align}
where $e$ is Euler's number.
\end{widetext}
A few comments are in order. Unlike in the last subsection, we did not have to take a large $\mathcal T$ limit since it takes only a finite amount of time to reach the worldsheet horizon. In fact, setting $\gamma$ to zero reduces the EE to $S_{EE}^{NC}(0)=-\frac{2}{3}\sqrt{\lambda}$ as we observed in the static case in~\cite{Grieninger:2023ehb}.
Moreover, the expression (\ref{SEE2}) is in general negative, 
or a loss due to radiation.

\subsection{Estimate of net  EE: strong field}
\begin{figure}
    \centering
    \includegraphics[width=1.0\linewidth]{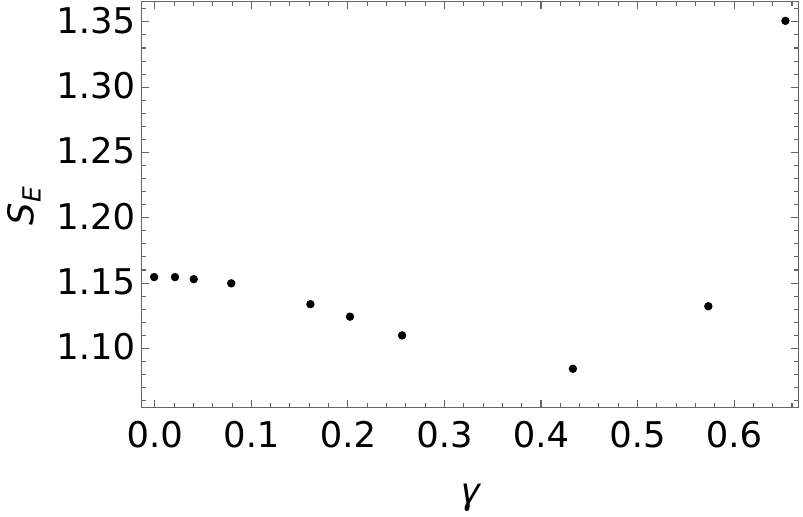} 
    \caption{ Net entanglement entropy for pair production in a pulse
    as in (\ref{SUGGESTED}) for $T=0.08$, $z_M=1$ and $\lambda=12$.}
    \label{fig:combined}
\end{figure}

The net EE in the pulse, is the sum of the causal contribution 
due to tunneling (positive) and the non-causal contribution due
to radiation (negative). In the strong field limit, the former follows from the Euclidean surface using (\ref{SEENUM}) as shown in Fig~\ref{fig:EEnumerics} (left). In contrast, the radiation part
following from the ruled world-sheet surface, only accounts for
the radiation in the weak-field limit (no back reaction) as we
argued in~\cite{Grieninger:2023ehb}. To remedy for this, we suggest 
an estimate for the net EE as 
\bea
\label{SUGGESTED}
S_{EE}=S^C_{EE}+\bigg(%\frac 23 
S^{NC}_{EE}+\sqrt\lambda \frac{T}{T_c}\bigg).
\eea
%where we have corrected for the overall factor of $\frac 23$ (see above). 
Note that with the added extra contribution, it reduces to the strong field limit discussed in~\cite{Grieninger:2023ehb} for $\omega=0$. As shown in Fig. \ref{fig:combined}, the radiation loss in (\ref{SUGGESTED})  slightly depletes the causal entanglement entropy for
intermediate $\gamma=M\omega/E$, before being overtaken by the latter for larger $\gamma$.
The dots in Fig. \ref{fig:combined} are the numerical results for (\ref{SUGGESTED}) with 
$S_{EE}^C$ following numerically from (\ref{SEENUM}),  and $S_{EE}^{NC}$
given by (\ref{SEE2}). %corrected by  $\frac 23$. 

\section{Conclusions}
\label{SECC}
In the semiclassical approximation at weak coupling, the Schwinger pair creation by an electric pulse is 
captured by a periodic worldline instanton. The inverse period of the instanton can be identified with the Unruh temperature.

The pair production in an electric  pulse at strong coupling in the gravity dual description is described by a worldsheet instanton. We have found the corresponding
worldsheet in Minkowski signature in the weak field limit. In the strong field limit, we have obtained numerically the tunneling surface, thereby generalizing the holographic Schwinger pair production process in a constant electric field~\cite{Semenoff:2011ng} to time-dependent electric pulses. 

Remarkably, the gravity dual string worldsheet exhibits a falling wormhole that acts as a separatrix splitting the worldsheet into a causal and acausal part which is hidden behind the horizon. In the weak field limit and under the assumption that $\omega$ is sufficiently large,  the causal
part of the worldsheet does not generate any entanglement entropy to leading order at late times, and all of it results from the radiation in the acausal part. 

This is not the case in the strong field limit, where the causal part of the worldsheet can be captured by the Euclidean surface. Indeed, the numerically generated Euclidean worldsheet results in a positive contribution to the  entanglement entropy. 

It would be interesting to derive the strong field expression for the contribution to the entanglement entropy from radiation. Moreover, in the weak field limit, it would be illuminating to work out the subleading contributions to the causal part of the entanglement entropy and see if it yields a net positive entanglement entropy at intermediate times.

\vskip 0.5cm
{\noindent\bf Acknowledgments}

\noindent 
This work was supported by the U.S. Department of Energy, Office of Science, Office of Nuclear Physics, Grants Nos. DE-FG88ER41450  and DE-SC0012704 (DK), and the U.S. Department of Energy, Office of Science, National Quantum Information Science Research Centers, Co-design Center for Quantum Advantage (C2QA) under Contract No.DE-SC0012704 (DK).

\bibliography{main}

%merlin.mbs apsrev4-1.bst 2010-07-25 4.21a (PWD, AO, DPC) hacked
%Control: key (0)
%Control: author (0) dotless jnrlst
%Control: editor formatted (1) identically to author
%Control: production of article title (0) allowed
%Control: page (1) range
%Control: year (0) verbatim
%Control: production of eprint (0) enabled
\begin{thebibliography}{35}%
\makeatletter
\providecommand \@ifxundefined [1]{%
 \@ifx{#1\undefined}
}%
\providecommand \@ifnum [1]{%
 \ifnum #1\expandafter \@firstoftwo
 \else \expandafter \@secondoftwo
 \fi
}%
\providecommand \@ifx [1]{%
 \ifx #1\expandafter \@firstoftwo
 \else \expandafter \@secondoftwo
 \fi
}%
\providecommand \natexlab [1]{#1}%
\providecommand \enquote  [1]{``#1''}%
\providecommand \bibnamefont  [1]{#1}%
\providecommand \bibfnamefont [1]{#1}%
\providecommand \citenamefont [1]{#1}%
\providecommand \href@noop [0]{\@secondoftwo}%
\providecommand \href [0]{\begingroup \@sanitize@url \@href}%
\providecommand \@href[1]{\@@startlink{#1}\@@href}%
\providecommand \@@href[1]{\endgroup#1\@@endlink}%
\providecommand \@sanitize@url [0]{\catcode `\\12\catcode `\$12\catcode
  `\&12\catcode `\#12\catcode `\^12\catcode `\_12\catcode `\%12\relax}%
\providecommand \@@startlink[1]{}%
\providecommand \@@endlink[0]{}%
\providecommand \url  [0]{\begingroup\@sanitize@url \@url }%
\providecommand \@url [1]{\endgroup\@href {#1}{\urlprefix }}%
\providecommand \urlprefix  [0]{URL }%
\providecommand \Eprint [0]{\href }%
\providecommand \doibase [0]{http://dx.doi.org/}%
\providecommand \selectlanguage [0]{\@gobble}%
\providecommand \bibinfo  [0]{\@secondoftwo}%
\providecommand \bibfield  [0]{\@secondoftwo}%
\providecommand \translation [1]{[#1]}%
\providecommand \BibitemOpen [0]{}%
\providecommand \bibitemStop [0]{}%
\providecommand \bibitemNoStop [0]{.\EOS\space}%
\providecommand \EOS [0]{\spacefactor3000\relax}%
\providecommand \BibitemShut  [1]{\csname bibitem#1\endcsname}%
\let\auto@bib@innerbib\@empty
%</preamble>
\bibitem [{\citenamefont {Schwinger}(1951)}]{Schwinger:1951nm}%
  \BibitemOpen
  \bibfield  {author} {\bibinfo {author} {\bibfnamefont {Julian~S.}\
  \bibnamefont {Schwinger}},\ }\bibfield  {title} {\enquote {\bibinfo {title}
  {{On gauge invariance and vacuum polarization}},}\ }\href {\doibase
  10.1103/PhysRev.82.664} {\bibfield  {journal} {\bibinfo  {journal} {Phys.
  Rev.}\ }\textbf {\bibinfo {volume} {82}},\ \bibinfo {pages} {664--679}
  (\bibinfo {year} {1951})}\BibitemShut {NoStop}%
\bibitem [{\citenamefont {Andersson}\ \emph {et~al.}(1983)\citenamefont
  {Andersson}, \citenamefont {Gustafson}, \citenamefont {Ingelman},\ and\
  \citenamefont {Sjostrand}}]{Andersson:1983ia}%
  \BibitemOpen
  \bibfield  {author} {\bibinfo {author} {\bibfnamefont {Bo}~\bibnamefont
  {Andersson}}, \bibinfo {author} {\bibfnamefont {G.}~\bibnamefont
  {Gustafson}}, \bibinfo {author} {\bibfnamefont {G.}~\bibnamefont {Ingelman}},
  \ and\ \bibinfo {author} {\bibfnamefont {T.}~\bibnamefont {Sjostrand}},\
  }\bibfield  {title} {\enquote {\bibinfo {title} {{Parton Fragmentation and
  String Dynamics}},}\ }\href {\doibase 10.1016/0370-1573(83)90080-7}
  {\bibfield  {journal} {\bibinfo  {journal} {Phys. Rept.}\ }\textbf {\bibinfo
  {volume} {97}},\ \bibinfo {pages} {31--145} (\bibinfo {year}
  {1983})}\BibitemShut {NoStop}%
\bibitem [{\citenamefont {Buskulic}\ \emph {et~al.}(1995)\citenamefont
  {Buskulic} \emph {et~al.}}]{ALEPH:1995qic}%
  \BibitemOpen
  \bibfield  {author} {\bibinfo {author} {\bibfnamefont {D.}~\bibnamefont
  {Buskulic}} \emph {et~al.} (\bibinfo {collaboration} {ALEPH}),\ }\bibfield
  {title} {\enquote {\bibinfo {title} {{Measurements of the charged particle
  multiplicity distribution in restricted rapidity intervals}},}\ }\href
  {\doibase 10.1007/BF02907382} {\bibfield  {journal} {\bibinfo  {journal} {Z.
  Phys. C}\ }\textbf {\bibinfo {volume} {69}},\ \bibinfo {pages} {15--26}
  (\bibinfo {year} {1995})}\BibitemShut {NoStop}%
\bibitem [{\citenamefont {Jensen}\ and\ \citenamefont
  {Karch}(2013)}]{Jensen:2013ora}%
  \BibitemOpen
  \bibfield  {author} {\bibinfo {author} {\bibfnamefont {Kristan}\ \bibnamefont
  {Jensen}}\ and\ \bibinfo {author} {\bibfnamefont {Andreas}\ \bibnamefont
  {Karch}},\ }\bibfield  {title} {\enquote {\bibinfo {title} {{Holographic Dual
  of an Einstein-Podolsky-Rosen Pair has a Wormhole}},}\ }\href {\doibase
  10.1103/PhysRevLett.111.211602} {\bibfield  {journal} {\bibinfo  {journal}
  {Phys. Rev. Lett.}\ }\textbf {\bibinfo {volume} {111}},\ \bibinfo {pages}
  {211602} (\bibinfo {year} {2013})},\ \Eprint {http://arxiv.org/abs/1307.1132}
  {arXiv:1307.1132 [hep-th]} \BibitemShut {NoStop}%
\bibitem [{\citenamefont {Sonner}(2013)}]{Sonner:2013mba}%
  \BibitemOpen
  \bibfield  {author} {\bibinfo {author} {\bibfnamefont {Julian}\ \bibnamefont
  {Sonner}},\ }\bibfield  {title} {\enquote {\bibinfo {title} {{Holographic
  Schwinger Effect and the Geometry of Entanglement}},}\ }\href {\doibase
  10.1103/PhysRevLett.111.211603} {\bibfield  {journal} {\bibinfo  {journal}
  {Phys. Rev. Lett.}\ }\textbf {\bibinfo {volume} {111}},\ \bibinfo {pages}
  {211603} (\bibinfo {year} {2013})},\ \Eprint {http://arxiv.org/abs/1307.6850}
  {arXiv:1307.6850 [hep-th]} \BibitemShut {NoStop}%
\bibitem [{\citenamefont {Gorsky}\ \emph {et~al.}(2002)\citenamefont {Gorsky},
  \citenamefont {Saraikin},\ and\ \citenamefont {Selivanov}}]{Gorsky:2001up}%
  \BibitemOpen
  \bibfield  {author} {\bibinfo {author} {\bibfnamefont {A.~S.}\ \bibnamefont
  {Gorsky}}, \bibinfo {author} {\bibfnamefont {K.~A.}\ \bibnamefont
  {Saraikin}}, \ and\ \bibinfo {author} {\bibfnamefont {K.~G.}\ \bibnamefont
  {Selivanov}},\ }\bibfield  {title} {\enquote {\bibinfo {title} {{Schwinger
  type processes via branes and their gravity duals}},}\ }\href {\doibase
  10.1016/S0550-3213(02)00095-0} {\bibfield  {journal} {\bibinfo  {journal}
  {Nucl. Phys. B}\ }\textbf {\bibinfo {volume} {628}},\ \bibinfo {pages}
  {270--294} (\bibinfo {year} {2002})},\ \Eprint
  {http://arxiv.org/abs/hep-th/0110178} {arXiv:hep-th/0110178} \BibitemShut
  {NoStop}%
\bibitem [{\citenamefont {Xiao}(2008)}]{Xiao:2008nr}%
  \BibitemOpen
  \bibfield  {author} {\bibinfo {author} {\bibfnamefont {Bo-Wen}\ \bibnamefont
  {Xiao}},\ }\bibfield  {title} {\enquote {\bibinfo {title} {{On the exact
  solution of the accelerating string in AdS(5) space}},}\ }\href {\doibase
  10.1016/j.physletb.2008.06.017} {\bibfield  {journal} {\bibinfo  {journal}
  {Phys. Lett. B}\ }\textbf {\bibinfo {volume} {665}},\ \bibinfo {pages}
  {173--177} (\bibinfo {year} {2008})},\ \Eprint
  {http://arxiv.org/abs/0804.1343} {arXiv:0804.1343 [hep-th]} \BibitemShut
  {NoStop}%
\bibitem [{\citenamefont {Semenoff}\ and\ \citenamefont
  {Zarembo}(2011)}]{Semenoff:2011ng}%
  \BibitemOpen
  \bibfield  {author} {\bibinfo {author} {\bibfnamefont {Gordon~W.}\
  \bibnamefont {Semenoff}}\ and\ \bibinfo {author} {\bibfnamefont {Konstantin}\
  \bibnamefont {Zarembo}},\ }\bibfield  {title} {\enquote {\bibinfo {title}
  {{Holographic Schwinger Effect}},}\ }\href {\doibase
  10.1103/PhysRevLett.107.171601} {\bibfield  {journal} {\bibinfo  {journal}
  {Phys. Rev. Lett.}\ }\textbf {\bibinfo {volume} {107}},\ \bibinfo {pages}
  {171601} (\bibinfo {year} {2011})},\ \Eprint {http://arxiv.org/abs/1109.2920}
  {arXiv:1109.2920 [hep-th]} \BibitemShut {NoStop}%
\bibitem [{\citenamefont {Lewkowycz}\ and\ \citenamefont
  {Maldacena}(2014)}]{Lewkowycz:2013laa}%
  \BibitemOpen
  \bibfield  {author} {\bibinfo {author} {\bibfnamefont {Aitor}\ \bibnamefont
  {Lewkowycz}}\ and\ \bibinfo {author} {\bibfnamefont {Juan}\ \bibnamefont
  {Maldacena}},\ }\bibfield  {title} {\enquote {\bibinfo {title} {{Exact
  results for the entanglement entropy and the energy radiated by a quark}},}\
  }\href {\doibase 10.1007/JHEP05(2014)025} {\bibfield  {journal} {\bibinfo
  {journal} {JHEP}\ }\textbf {\bibinfo {volume} {05}},\ \bibinfo {pages} {025}
  (\bibinfo {year} {2014})},\ \Eprint {http://arxiv.org/abs/1312.5682}
  {arXiv:1312.5682 [hep-th]} \BibitemShut {NoStop}%
\bibitem [{\citenamefont {Chernicoff}\ \emph {et~al.}(2013)\citenamefont
  {Chernicoff}, \citenamefont {G\"uijosa},\ and\ \citenamefont
  {Pedraza}}]{Chernicoff:2013iga}%
  \BibitemOpen
  \bibfield  {author} {\bibinfo {author} {\bibfnamefont {Mariano}\ \bibnamefont
  {Chernicoff}}, \bibinfo {author} {\bibfnamefont {Alberto}\ \bibnamefont
  {G\"uijosa}}, \ and\ \bibinfo {author} {\bibfnamefont {Juan~F.}\ \bibnamefont
  {Pedraza}},\ }\bibfield  {title} {\enquote {\bibinfo {title} {{Holographic
  EPR Pairs, Wormholes and Radiation}},}\ }\href {\doibase
  10.1007/JHEP10(2013)211} {\bibfield  {journal} {\bibinfo  {journal} {JHEP}\
  }\textbf {\bibinfo {volume} {10}},\ \bibinfo {pages} {211} (\bibinfo {year}
  {2013})},\ \Eprint {http://arxiv.org/abs/1308.3695} {arXiv:1308.3695
  [hep-th]} \BibitemShut {NoStop}%
\bibitem [{\citenamefont {Jensen}\ \emph {et~al.}(2014)\citenamefont {Jensen},
  \citenamefont {Karch},\ and\ \citenamefont {Robinson}}]{Jensen:2014bpa}%
  \BibitemOpen
  \bibfield  {author} {\bibinfo {author} {\bibfnamefont {Kristan}\ \bibnamefont
  {Jensen}}, \bibinfo {author} {\bibfnamefont {Andreas}\ \bibnamefont {Karch}},
  \ and\ \bibinfo {author} {\bibfnamefont {Brandon}\ \bibnamefont {Robinson}},\
  }\bibfield  {title} {\enquote {\bibinfo {title} {{Holographic dual of a
  Hawking pair has a wormhole}},}\ }\href {\doibase 10.1103/PhysRevD.90.064019}
  {\bibfield  {journal} {\bibinfo  {journal} {Phys. Rev. D}\ }\textbf {\bibinfo
  {volume} {90}},\ \bibinfo {pages} {064019} (\bibinfo {year} {2014})},\
  \Eprint {http://arxiv.org/abs/1405.2065} {arXiv:1405.2065 [hep-th]}
  \BibitemShut {NoStop}%
\bibitem [{\citenamefont {Hubeny}\ and\ \citenamefont
  {Semenoff}(2014)}]{Hubeny:2014zna}%
  \BibitemOpen
  \bibfield  {author} {\bibinfo {author} {\bibfnamefont {Veronika~E.}\
  \bibnamefont {Hubeny}}\ and\ \bibinfo {author} {\bibfnamefont {Gordon~W.}\
  \bibnamefont {Semenoff}},\ }\bibfield  {title} {\enquote {\bibinfo {title}
  {{Holographic Accelerated Heavy Quark-Anti-Quark Pair}},}\ }\href@noop {} {\
  (\bibinfo {year} {2014})},\ \Eprint {http://arxiv.org/abs/1410.1172}
  {arXiv:1410.1172 [hep-th]} \BibitemShut {NoStop}%
\bibitem [{\citenamefont {Ghodrati}(2015)}]{Ghodrati:2015rta}%
  \BibitemOpen
  \bibfield  {author} {\bibinfo {author} {\bibfnamefont {Mahdis}\ \bibnamefont
  {Ghodrati}},\ }\bibfield  {title} {\enquote {\bibinfo {title} {{Schwinger
  Effect and Entanglement Entropy in Confining Geometries}},}\ }\href {\doibase
  10.1103/PhysRevD.92.065015} {\bibfield  {journal} {\bibinfo  {journal} {Phys.
  Rev. D}\ }\textbf {\bibinfo {volume} {92}},\ \bibinfo {pages} {065015}
  (\bibinfo {year} {2015})},\ \Eprint {http://arxiv.org/abs/1506.08557}
  {arXiv:1506.08557 [hep-th]} \BibitemShut {NoStop}%
\bibitem [{\citenamefont {Semenoff}(2018)}]{Semenoff:2018ffq}%
  \BibitemOpen
  \bibfield  {author} {\bibinfo {author} {\bibfnamefont {Gordon~W.}\
  \bibnamefont {Semenoff}},\ }\bibfield  {title} {\enquote {\bibinfo {title}
  {{Lectures on the holographic duality of gauge fields and strings}},}\ }\href
  {\doibase 10.1093/oso/9780198828150.003.0003} {\  (\bibinfo {year} {2018}),\
  10.1093/oso/9780198828150.003.0003},\ \Eprint
  {http://arxiv.org/abs/1808.04074} {arXiv:1808.04074 [hep-th]} \BibitemShut
  {NoStop}%
\bibitem [{\citenamefont {Yeh}(2023)}]{Yeh:2023avs}%
  \BibitemOpen
  \bibfield  {author} {\bibinfo {author} {\bibfnamefont {Chen-Pin}\
  \bibnamefont {Yeh}},\ }\bibfield  {title} {\enquote {\bibinfo {title} {{Shock
  Waves in Holographic EPR pair}},}\ }\href@noop {} {\  (\bibinfo {year}
  {2023})},\ \Eprint {http://arxiv.org/abs/2310.00991} {arXiv:2310.00991
  [hep-th]} \BibitemShut {NoStop}%
\bibitem [{\citenamefont {Maldacena}\ and\ \citenamefont
  {Susskind}(2013)}]{Maldacena:2013xja}%
  \BibitemOpen
  \bibfield  {author} {\bibinfo {author} {\bibfnamefont {Juan}\ \bibnamefont
  {Maldacena}}\ and\ \bibinfo {author} {\bibfnamefont {Leonard}\ \bibnamefont
  {Susskind}},\ }\bibfield  {title} {\enquote {\bibinfo {title} {{Cool horizons
  for entangled black holes}},}\ }\href {\doibase 10.1002/prop.201300020}
  {\bibfield  {journal} {\bibinfo  {journal} {Fortsch. Phys.}\ }\textbf
  {\bibinfo {volume} {61}},\ \bibinfo {pages} {781--811} (\bibinfo {year}
  {2013})},\ \Eprint {http://arxiv.org/abs/1306.0533} {arXiv:1306.0533
  [hep-th]} \BibitemShut {NoStop}%
\bibitem [{\citenamefont {Grieninger}\ \emph {et~al.}(2023)\citenamefont
  {Grieninger}, \citenamefont {Kharzeev},\ and\ \citenamefont
  {Zahed}}]{Grieninger:2023ehb}%
  \BibitemOpen
  \bibfield  {author} {\bibinfo {author} {\bibfnamefont {Sebastian}\
  \bibnamefont {Grieninger}}, \bibinfo {author} {\bibfnamefont {Dmitri~E.}\
  \bibnamefont {Kharzeev}}, \ and\ \bibinfo {author} {\bibfnamefont {Ismail}\
  \bibnamefont {Zahed}},\ }\bibfield  {title} {\enquote {\bibinfo {title}
  {{Entanglement in a holographic Schwinger pair with confinement}},}\ }\href
  {\doibase 10.1103/PhysRevD.108.086030} {\bibfield  {journal} {\bibinfo
  {journal} {Phys. Rev. D}\ }\textbf {\bibinfo {volume} {108}},\ \bibinfo
  {pages} {086030} (\bibinfo {year} {2023})},\ \Eprint
  {http://arxiv.org/abs/2305.07121} {arXiv:2305.07121 [hep-th]} \BibitemShut
  {NoStop}%
\bibitem [{\citenamefont {Hubeny}\ and\ \citenamefont
  {Semenoff}(2015)}]{Hubeny:2014kma}%
  \BibitemOpen
  \bibfield  {author} {\bibinfo {author} {\bibfnamefont {Veronika~E.}\
  \bibnamefont {Hubeny}}\ and\ \bibinfo {author} {\bibfnamefont {Gordon~W.}\
  \bibnamefont {Semenoff}},\ }\bibfield  {title} {\enquote {\bibinfo {title}
  {{String worldsheet for accelerating quark}},}\ }\href {\doibase
  10.1007/JHEP10(2015)071} {\bibfield  {journal} {\bibinfo  {journal} {JHEP}\
  }\textbf {\bibinfo {volume} {10}},\ \bibinfo {pages} {071} (\bibinfo {year}
  {2015})},\ \Eprint {http://arxiv.org/abs/1410.1171} {arXiv:1410.1171
  [hep-th]} \BibitemShut {NoStop}%
\bibitem [{\citenamefont {Gelis}\ \emph {et~al.}(2010)\citenamefont {Gelis},
  \citenamefont {Iancu}, \citenamefont {Jalilian-Marian},\ and\ \citenamefont
  {Venugopalan}}]{Gelis:2010nm}%
  \BibitemOpen
  \bibfield  {author} {\bibinfo {author} {\bibfnamefont {Francois}\
  \bibnamefont {Gelis}}, \bibinfo {author} {\bibfnamefont {Edmond}\
  \bibnamefont {Iancu}}, \bibinfo {author} {\bibfnamefont {Jamal}\ \bibnamefont
  {Jalilian-Marian}}, \ and\ \bibinfo {author} {\bibfnamefont {Raju}\
  \bibnamefont {Venugopalan}},\ }\bibfield  {title} {\enquote {\bibinfo {title}
  {{The Color Glass Condensate}},}\ }\href {\doibase
  10.1146/annurev.nucl.010909.083629} {\bibfield  {journal} {\bibinfo
  {journal} {Ann. Rev. Nucl. Part. Sci.}\ }\textbf {\bibinfo {volume} {60}},\
  \bibinfo {pages} {463--489} (\bibinfo {year} {2010})},\ \Eprint
  {http://arxiv.org/abs/1002.0333} {arXiv:1002.0333 [hep-ph]} \BibitemShut
  {NoStop}%
\bibitem [{\citenamefont {Kovchegov}\ and\ \citenamefont
  {Levin}(2013)}]{Kovchegov:2012mbw}%
  \BibitemOpen
  \bibfield  {author} {\bibinfo {author} {\bibfnamefont {Yuri~V.}\ \bibnamefont
  {Kovchegov}}\ and\ \bibinfo {author} {\bibfnamefont {Eugene}\ \bibnamefont
  {Levin}},\ }\href {\doibase 10.1017/9781009291446} {\emph {\bibinfo {title}
  {{Quantum Chromodynamics at High Energy}}}},\ Vol.~\bibinfo {volume} {33}\
  (\bibinfo  {publisher} {Oxford University Press},\ \bibinfo {year}
  {2013})\BibitemShut {NoStop}%
\bibitem [{\citenamefont {Bjorken}(1992)}]{bjorken1992black}%
  \BibitemOpen
  \bibfield  {author} {\bibinfo {author} {\bibfnamefont {James~D}\ \bibnamefont
  {Bjorken}},\ }\bibfield  {title} {\enquote {\bibinfo {title} {How black is a
  constituent quark??}}\ }\href@noop {} {\bibfield  {journal} {\bibinfo
  {journal} {Acta Physica Polonica B}\ }\textbf {\bibinfo {volume} {23}},\
  \bibinfo {pages} {637} (\bibinfo {year} {1992})}\BibitemShut {NoStop}%
\bibitem [{\citenamefont {Florio}\ and\ \citenamefont
  {Kharzeev}(2021)}]{florio2021gibbs}%
  \BibitemOpen
  \bibfield  {author} {\bibinfo {author} {\bibfnamefont {Adrien}\ \bibnamefont
  {Florio}}\ and\ \bibinfo {author} {\bibfnamefont {Dmitri~E}\ \bibnamefont
  {Kharzeev}},\ }\bibfield  {title} {\enquote {\bibinfo {title} {Gibbs entropy
  from entanglement in electric quenches},}\ }\href@noop {} {\bibfield
  {journal} {\bibinfo  {journal} {Physical Review D}\ }\textbf {\bibinfo
  {volume} {104}},\ \bibinfo {pages} {056021} (\bibinfo {year}
  {2021})}\BibitemShut {NoStop}%
\bibitem [{\citenamefont {Casini}\ \emph {et~al.}(2011)\citenamefont {Casini},
  \citenamefont {Huerta},\ and\ \citenamefont {Myers}}]{Casini:2011kv}%
  \BibitemOpen
  \bibfield  {author} {\bibinfo {author} {\bibfnamefont {Horacio}\ \bibnamefont
  {Casini}}, \bibinfo {author} {\bibfnamefont {Marina}\ \bibnamefont {Huerta}},
  \ and\ \bibinfo {author} {\bibfnamefont {Robert~C.}\ \bibnamefont {Myers}},\
  }\bibfield  {title} {\enquote {\bibinfo {title} {{Towards a derivation of
  holographic entanglement entropy}},}\ }\href {\doibase
  10.1007/JHEP05(2011)036} {\bibfield  {journal} {\bibinfo  {journal} {JHEP}\
  }\textbf {\bibinfo {volume} {05}},\ \bibinfo {pages} {036} (\bibinfo {year}
  {2011})},\ \Eprint {http://arxiv.org/abs/1102.0440} {arXiv:1102.0440
  [hep-th]} \BibitemShut {NoStop}%
\bibitem [{\citenamefont {Brezin}\ and\ \citenamefont
  {Itzykson}(1970)}]{Brezin:1970xf}%
  \BibitemOpen
  \bibfield  {author} {\bibinfo {author} {\bibfnamefont {E.}~\bibnamefont
  {Brezin}}\ and\ \bibinfo {author} {\bibfnamefont {C.}~\bibnamefont
  {Itzykson}},\ }\bibfield  {title} {\enquote {\bibinfo {title} {{Pair
  production in vacuum by an alternating field}},}\ }\href {\doibase
  10.1103/PhysRevD.2.1191} {\bibfield  {journal} {\bibinfo  {journal} {Phys.
  Rev. D}\ }\textbf {\bibinfo {volume} {2}},\ \bibinfo {pages} {1191--1199}
  (\bibinfo {year} {1970})}\BibitemShut {NoStop}%
\bibitem [{\citenamefont {Popov}(1971)}]{Popov:1971iga}%
  \BibitemOpen
  \bibfield  {author} {\bibinfo {author} {\bibfnamefont {V.~S.}\ \bibnamefont
  {Popov}},\ }\bibfield  {title} {\enquote {\bibinfo {title} {{Pair production
  in a variable external field (quasiclassical approximation)}},}\ }\href@noop
  {} {\bibfield  {journal} {\bibinfo  {journal} {Zh. Eksp. Teor. Fiz.}\
  }\textbf {\bibinfo {volume} {61}},\ \bibinfo {pages} {1334--1351} (\bibinfo
  {year} {1971})}\BibitemShut {NoStop}%
\bibitem [{\citenamefont {Kim}\ and\ \citenamefont {Page}(2002)}]{Kim:2000un}%
  \BibitemOpen
  \bibfield  {author} {\bibinfo {author} {\bibfnamefont {Sang~Pyo}\
  \bibnamefont {Kim}}\ and\ \bibinfo {author} {\bibfnamefont {Don~N.}\
  \bibnamefont {Page}},\ }\bibfield  {title} {\enquote {\bibinfo {title}
  {{Schwinger pair production via instantons in a strong electric field}},}\
  }\href {\doibase 10.1103/PhysRevD.65.105002} {\bibfield  {journal} {\bibinfo
  {journal} {Phys. Rev. D}\ }\textbf {\bibinfo {volume} {65}},\ \bibinfo
  {pages} {105002} (\bibinfo {year} {2002})},\ \Eprint
  {http://arxiv.org/abs/hep-th/0005078} {arXiv:hep-th/0005078} \BibitemShut
  {NoStop}%
\bibitem [{\citenamefont {Dunne}\ and\ \citenamefont
  {Schubert}(2005)}]{Dunne:2005sx}%
  \BibitemOpen
  \bibfield  {author} {\bibinfo {author} {\bibfnamefont {Gerald~V.}\
  \bibnamefont {Dunne}}\ and\ \bibinfo {author} {\bibfnamefont {Christian}\
  \bibnamefont {Schubert}},\ }\bibfield  {title} {\enquote {\bibinfo {title}
  {{Worldline instantons and pair production in inhomogeneous fields}},}\
  }\href {\doibase 10.1103/PhysRevD.72.105004} {\bibfield  {journal} {\bibinfo
  {journal} {Phys. Rev. D}\ }\textbf {\bibinfo {volume} {72}},\ \bibinfo
  {pages} {105004} (\bibinfo {year} {2005})},\ \Eprint
  {http://arxiv.org/abs/hep-th/0507174} {arXiv:hep-th/0507174} \BibitemShut
  {NoStop}%
\bibitem [{\citenamefont {Kharzeev}\ and\ \citenamefont
  {Tuchin}(2005)}]{kharzeev2005color}%
  \BibitemOpen
  \bibfield  {author} {\bibinfo {author} {\bibfnamefont {Dmitri}\ \bibnamefont
  {Kharzeev}}\ and\ \bibinfo {author} {\bibfnamefont {Kirill}\ \bibnamefont
  {Tuchin}},\ }\bibfield  {title} {\enquote {\bibinfo {title} {From color glass
  condensate to quark--gluon plasma through the event horizon},}\ }\href@noop
  {} {\bibfield  {journal} {\bibinfo  {journal} {Nuclear Physics A}\ }\textbf
  {\bibinfo {volume} {753}},\ \bibinfo {pages} {316--334} (\bibinfo {year}
  {2005})}\BibitemShut {NoStop}%
\bibitem [{\citenamefont {Gies}\ and\ \citenamefont
  {Klingmuller}(2005)}]{Gies:2005bz}%
  \BibitemOpen
  \bibfield  {author} {\bibinfo {author} {\bibfnamefont {Holger}\ \bibnamefont
  {Gies}}\ and\ \bibinfo {author} {\bibfnamefont {Klaus}\ \bibnamefont
  {Klingmuller}},\ }\bibfield  {title} {\enquote {\bibinfo {title} {{Pair
  production in inhomogeneous fields}},}\ }\href {\doibase
  10.1103/PhysRevD.72.065001} {\bibfield  {journal} {\bibinfo  {journal} {Phys.
  Rev. D}\ }\textbf {\bibinfo {volume} {72}},\ \bibinfo {pages} {065001}
  (\bibinfo {year} {2005})},\ \Eprint {http://arxiv.org/abs/hep-ph/0505099}
  {arXiv:hep-ph/0505099} \BibitemShut {NoStop}%
\bibitem [{\citenamefont {Gies}\ and\ \citenamefont
  {Torgrimsson}(2016)}]{Gies:2015hia}%
  \BibitemOpen
  \bibfield  {author} {\bibinfo {author} {\bibfnamefont {Holger}\ \bibnamefont
  {Gies}}\ and\ \bibinfo {author} {\bibfnamefont {Greger}\ \bibnamefont
  {Torgrimsson}},\ }\bibfield  {title} {\enquote {\bibinfo {title} {{Critical
  Schwinger pair production}},}\ }\href {\doibase
  10.1103/PhysRevLett.116.090406} {\bibfield  {journal} {\bibinfo  {journal}
  {Phys. Rev. Lett.}\ }\textbf {\bibinfo {volume} {116}},\ \bibinfo {pages}
  {090406} (\bibinfo {year} {2016})},\ \Eprint
  {http://arxiv.org/abs/1507.07802} {arXiv:1507.07802 [hep-ph]} \BibitemShut
  {NoStop}%
\bibitem [{\citenamefont {Gies}\ and\ \citenamefont
  {Torgrimsson}(2017)}]{Gies:2016coz}%
  \BibitemOpen
  \bibfield  {author} {\bibinfo {author} {\bibfnamefont {Holger}\ \bibnamefont
  {Gies}}\ and\ \bibinfo {author} {\bibfnamefont {Greger}\ \bibnamefont
  {Torgrimsson}},\ }\bibfield  {title} {\enquote {\bibinfo {title} {{Critical
  Schwinger pair production II - universality in the deeply critical
  regime}},}\ }\href {\doibase 10.1103/PhysRevD.95.016001} {\bibfield
  {journal} {\bibinfo  {journal} {Phys. Rev. D}\ }\textbf {\bibinfo {volume}
  {95}},\ \bibinfo {pages} {016001} (\bibinfo {year} {2017})},\ \Eprint
  {http://arxiv.org/abs/1612.00635} {arXiv:1612.00635 [hep-th]} \BibitemShut
  {NoStop}%
\bibitem [{\citenamefont {Mikhailov}(2003)}]{Mikhailov:2003er}%
  \BibitemOpen
  \bibfield  {author} {\bibinfo {author} {\bibfnamefont {Andrei}\ \bibnamefont
  {Mikhailov}},\ }\bibfield  {title} {\enquote {\bibinfo {title} {{Nonlinear
  waves in AdS / CFT correspondence}},}\ }\href@noop {} {\  (\bibinfo {year}
  {2003})},\ \Eprint {http://arxiv.org/abs/hep-th/0305196}
  {arXiv:hep-th/0305196} \BibitemShut {NoStop}%
\bibitem [{\citenamefont {Shuryak}\ \emph {et~al.}(2007)\citenamefont
  {Shuryak}, \citenamefont {Sin},\ and\ \citenamefont
  {Zahed}}]{Shuryak:2005ia}%
  \BibitemOpen
  \bibfield  {author} {\bibinfo {author} {\bibfnamefont {Edward}\ \bibnamefont
  {Shuryak}}, \bibinfo {author} {\bibfnamefont {Sang-Jin}\ \bibnamefont {Sin}},
  \ and\ \bibinfo {author} {\bibfnamefont {Ismail}\ \bibnamefont {Zahed}},\
  }\bibfield  {title} {\enquote {\bibinfo {title} {{A Gravity dual of RHIC
  collisions}},}\ }\href {\doibase 10.3938/jkps.50.384} {\bibfield  {journal}
  {\bibinfo  {journal} {J. Korean Phys. Soc.}\ }\textbf {\bibinfo {volume}
  {50}},\ \bibinfo {pages} {384--397} (\bibinfo {year} {2007})},\ \Eprint
  {http://arxiv.org/abs/hep-th/0511199} {arXiv:hep-th/0511199} \BibitemShut
  {NoStop}%
\bibitem [{\citenamefont {Kim}\ \emph {et~al.}(2008)\citenamefont {Kim},
  \citenamefont {Sin},\ and\ \citenamefont {Zahed}}]{Kim:2007ut}%
  \BibitemOpen
  \bibfield  {author} {\bibinfo {author} {\bibfnamefont {Keun-Young}\
  \bibnamefont {Kim}}, \bibinfo {author} {\bibfnamefont {Sang-Jin}\
  \bibnamefont {Sin}}, \ and\ \bibinfo {author} {\bibfnamefont {Ismail}\
  \bibnamefont {Zahed}},\ }\bibfield  {title} {\enquote {\bibinfo {title}
  {{Diffusion in an expanding plasma using AdS/CFT}},}\ }\href {\doibase
  10.1088/1126-6708/2008/04/047} {\bibfield  {journal} {\bibinfo  {journal}
  {JHEP}\ }\textbf {\bibinfo {volume} {04}},\ \bibinfo {pages} {047} (\bibinfo
  {year} {2008})},\ \Eprint {http://arxiv.org/abs/0707.0601} {arXiv:0707.0601
  [hep-th]} \BibitemShut {NoStop}%
\bibitem [{\citenamefont {Grieninger}\ and\ \citenamefont
  {Zahed}(2023)}]{Grieninger:2022yps}%
  \BibitemOpen
  \bibfield  {author} {\bibinfo {author} {\bibfnamefont {Sebastian}\
  \bibnamefont {Grieninger}}\ and\ \bibinfo {author} {\bibfnamefont {Ismail}\
  \bibnamefont {Zahed}},\ }\bibfield  {title} {\enquote {\bibinfo {title}
  {{Out-of-equilibrium photon production and electric conductivity in a
  holographic Bjorken expanding plasma}},}\ }\href {\doibase
  10.1103/PhysRevD.107.046017} {\bibfield  {journal} {\bibinfo  {journal}
  {Phys. Rev. D}\ }\textbf {\bibinfo {volume} {107}},\ \bibinfo {pages}
  {046017} (\bibinfo {year} {2023})},\ \Eprint
  {http://arxiv.org/abs/2211.10372} {arXiv:2211.10372 [hep-ph]} \BibitemShut
  {NoStop}%
\end{thebibliography}%
\end{document}